# Two-fold anisotropic superconducting state in topological superconductor $Sn_4Au$

M.M. SHARMA[1,2], GANESH GURJAR[3,4], S. PATNAIK[3] and V.P.S. AWANA[1,2,*]

[1]Academy of Scientific and Innovative Research (AcSIR), Ghaziabad-201002, India
[2]CSIR- National Physical Laboratory, New Delhi-110012, India
[3]School of Physical Sciences, Jawaharlal Nehru University, New Delhi-110067, India
[4]Ramjas College, New Delhi-110007, India

*E-mail: awana@nplindia.org: Home page: awanavps.webs.com

**Abstract** – Here we report the anisotropic magnetotransport properties in the superconducting state of $Sn_4Au$ single crystal. $Sn_4Au$ single crystal is synthesized through an easy melt growth method. Superconducting properties are evidenced from resistivity vs. temperature ($\rho$-T) and DC magnetization measurements. Isothermal magnetization measurements (M-H) hint toward type-II superconductivity in $Sn_4Au$. In-plane and out-of-plane $\rho$-H measurements show anisotropic behavior of the upper critical field at temperatures below superconducting transition ($T_c$ = 2.3 K). The observed anisotropy is more elucidated in $\rho$-H measurements performed below $T_c$ at different tilt angles. The anisotropy parameter ($\Gamma$) is found to be 1.26. The observed results show the presence two-fold anisotropic superconducting state in $Sn_4Au$ single crystal, which may be induced due to the layered structure of synthesized $Sn_4Au$ single crystal.

**Introduction**– Superconducting topological materials are envisaged to exhibit a plethora of interesting properties such as the existence of Majorana fermions on their boundary, breaking of crystalline symmetry in the superconducting state, and unconventional superconducting gap, etc [1-6]. In topological superconductors, the bulk band gap in topological insulators (TIs) is replaced by a bulk superconducting gap [5]. Topological superconductivity in bulk materials was first observed in doped TIs viz. $A_xBi_2Se_3$ (A=Nb, Sr, Cu) [7-11], $A_xBi_2Te_3$ (A=Pd, Tl) [12,13], and $In_xSn_{1-x}Te$ [14]. Some of the Topological semimetals (TSMs) viz. $PdTe_2$ [15], $CaSn_3$ [16], $CaSb_2$ [17], $Sn_4Au$ [18], YPtBi [19], etc, are also found to show the simultaneous existence of topological surface states and bulk superconductivity. Interestingly, the doped TIs show broken rotational symmetry in the superconducting state leading to a nematic superconducting phase [20-22]. Most topological superconductors are reported to show two-fold anisotropy in their superconducting state, resulting from spontaneously broken rotational symmetry [20-24]. Breaking of crystalline symmetry in a superconducting state can be probed through μ-SR measurements, nuclear magnetic resonance (NMR) measurements, and angle-dependent magnetotransport measurements [20-24]. In-plane anisotropy in the superconducting upper critical field provides evidence of the presence of two-fold symmetry in the superconducting state of the studied material [20-23].

Recently, $Sn_4Au$ crystallizing with an orthorhombic crystal structure has emerged as a suitable candidate for topological superconductivity [18,25-27]. $Sn_4Au$ undergoes a superconducting transition below 2.4 K, showing a significant magnetoresistance (MR) in its normal state [18,25,26]. The normal state of $Sn_4Au$ is reported to have a weak antilocalization effect emerging due to the presence of topological surface states [25]. Theoretical calculations on $Sn_4Au$ suggest the same to have a bulk electronic band structure resembling to high symmetry line semimetals (HLSMs) [27]. The recent reports on two-fold crystalline symmetry in the superconducting state of topological semimetals and doped TIs, inspire us to check the same for another superconducting TSM, $Sn_4Au$. In this letter, we report a detailed study of the superconducting state of synthesized $Sn_4Au$ single crystal. $Sn_4Au$ crystal is synthesized following a simple melt growth method. The as-grown crystal undergoes the superconducting transition below 2.3 K, as evidenced by resistivity vs. temperature ($\rho$-T) and DC magnetization measurements. The upper critical field ($H_{c2}$) is measured at different orientations of the magnetic field with respect to the basal plane, and the same is found to show unusual anisotropic behavior. This anisotropy in $H_{c2}$ is similar to as observed for other layered topological superconducting materials showing breaking of rotational symmetry in the superconducting phase. This short letter provides magneto transport evidence of a two-fold anisotropic superconducting phase in $Sn_4Au$, which may have eventually emerged from the layered non-centrosymmetric structure of the same.

**Experimental:** Single crystalline $Sn_4Au$ has been synthesized using a simple melt growth method. The details of heat treatment followed in the synthesis of $Sn_4Au$ single crystals have been reported earlier by us [25]. Rigaku mini flex II tabletop XRD equipped with Cu $K_\alpha$ radiation has been used to record XRD pattern on mechanically cleaved crystal flake and powdered sample. Full Prof software is used for Rietveld refinement of the PXRD pattern. The crystallographic information file (CIF) generated from Rietveld refinement is processed in VESTA software to draw the unit cells of $Sn_4Au$ single crystal. Magnetotransport measurements have been performed using the conventional four-probe method on Quantum Design Physical Property Measurement System (QD-PPMS) equipped with a sample horizontal rotator.

**Results & Discussion:** Fig. 1(a) shows the Rietveld refined powder XRD (PXRD) pattern taken on the gently crushed powder of synthesized $Sn_4Au$ single crystal. $Sn_4Au$ is reported to be crystallized with one of the two space group symmetries, Aea2 [18,28] and Ccce [29]. Both of the space groups are indistinguishable within the limit of XRD as they have the same Bragg positions. Here, Rietveld refinement is performed considering both of the space groups. The upper panel of fig. 1(a) shows the Rietveld refined PXRD pattern considering the Ccce space group symmetry, and the lower one shows the same with Aea2 space group symmetry parameters. Bragg's positions seemed to be the same for both of the space groups; the only difference can be made in terms of the intensity of XRD peaks. Intensities are seemed to be well-fitted for Aea2 space group rather than the Ccce one. The results of Rietveld refinement are given in Table 1. According to the fitting parameters obtained from the Rietveld refinement, XRD peaks are more fitted in the Aea2 space group. The previous experimental
reports on $Sn_4Au$ [18,25,28] also suggest the same to be crystallized in Aea2 space group symmetry.

Unit cells of $Sn_4Au$ have been drawn for the two space groups, and the same are shown as the insets in fig. 1(a). A clear centrosymmetric is visible in the unit cell drawn with the Ccce space group, which is absent in Aea2. Also, Sn has two independent sites, Sn1 and Sn2, in the Aea2 space group, which are transformed into a single atomic site, i.e., Sn in the Ccce space group. $Sn_4Au$ has a layered structure in both of the space groups. Stacking of the layers in the Ccce space group occurs along the b-axis, while for the Aea2 space group, stacking occurs along the c-axis.

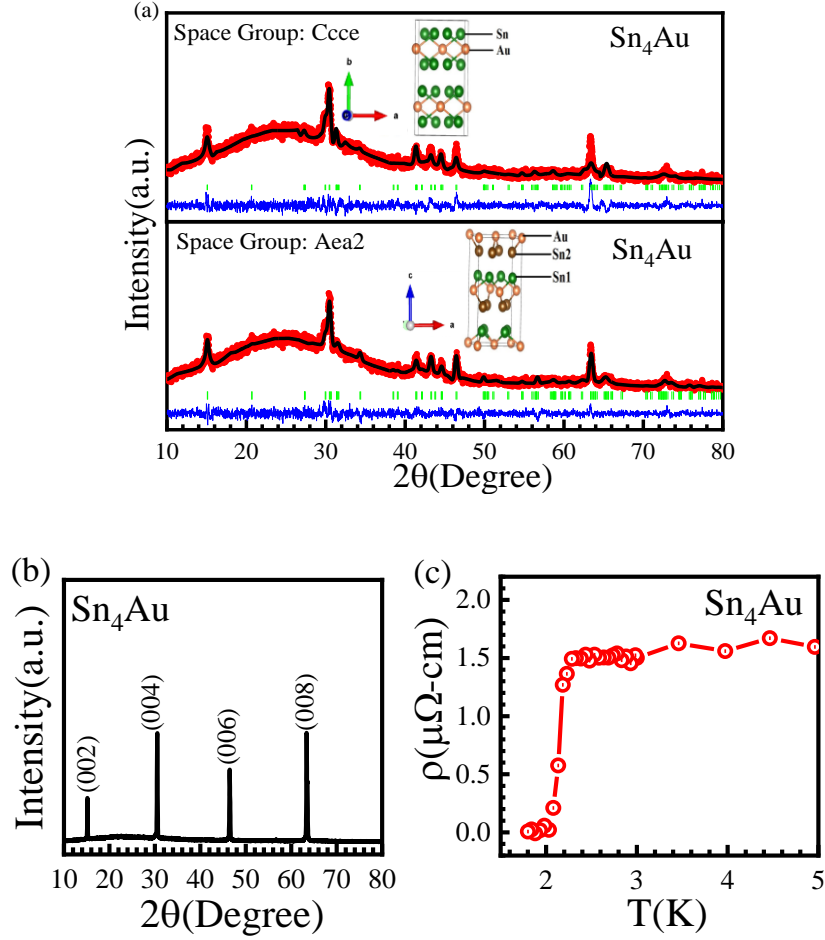

**Fig. 1: (a)** Rietveld refined PXRD pattern of synthesized Sn₄Au single crystal in which the red symbols are showing the experimental XRD data while the solid black line is showing the fitted XRD plot, the upper and lower panels are showing the fitted PXRD pattern along with VESTA drawn unit cells in Ccce and Aea2 space groups respectively. **(b)** XRD pattern taken mechanically cleaved crystal flake confirming the unidirectional growth of the crystal. **(c)** ρ-T measurements plot in the proximity of superconducting transition.

**Table-1**

Parameters obtained from Rietveld refinement:

| Space group: Aea2 (41) | | Space Group: Ccce (68) | |
|---|---|---|---|
| Cell Parameters | Refinement Parameters | Cell Parameters | Refinement Parameters |
| Cell type: Orthorhombic<br>Space Group: Aea2 (41)<br>Lattice parameters: a=6.525(1)Å<br>b=6.521(6)Å & c=11.731(7) Å<br>α=β=γ=90°<br>Cell volume: 499.148(4)Å³<br>Density: 8.937(3) g/cm³<br>Atomic co-ordinates:<br>Sn1 (0.1694,0.3395,0.1242)<br>Au (0,0,0)<br>Sn2 (0.3501,0.1624,0.8573) | $\chi^2$=1.36<br>$R_p$=5.20<br>$R_{wp}$=6.76<br>$R_{exp}$=5.79 | Cell type: Orthorhombic<br>Space Group: Ccce (68)<br>Lattice parameters: a=6.547(1)Å<br>b=11.756(6)Å & c=6.496(0) Å<br>α=β=γ=90°<br>Cell volume: 499.974(2)Å³<br>Density: 7.904(6) g/cm³<br>Atomic co-ordinates:<br>Sn (0.3476,0.1222,0.2176)<br>Au (0,0.25,0.25) | $\chi^2$=1.92<br>$R_p$=9.25<br>$R_{wp}$=12.1<br>$R_{exp}$=8.61 |

Fig. 1(b) shows the XRD pattern taken on mechanically cleaved crystal flake and depicts the high-intensity XRD peaks at some specific 2θ values, which corresponds to reflections from the (002n) planes for Aea2 space group and (02n0) for the Ccce one. Here, in the present article, the synthesized $Sn_4Au$ crystals are considered to be crystallized with Aea2 space group symmetry, so only (002n) planes, i.e., (002), (004), (006) and (008) are mentioned in the fig. 1(b). This shows that the synthesized $Sn_4Au$ single crystal is grown unidirectionally and confirms the crystallinity of the same. The full width at half maxima (FWHM) of the highest intensity peak is 0.12°, showing the crystalline character of the synthesized $Sn_4Au$ crystal.

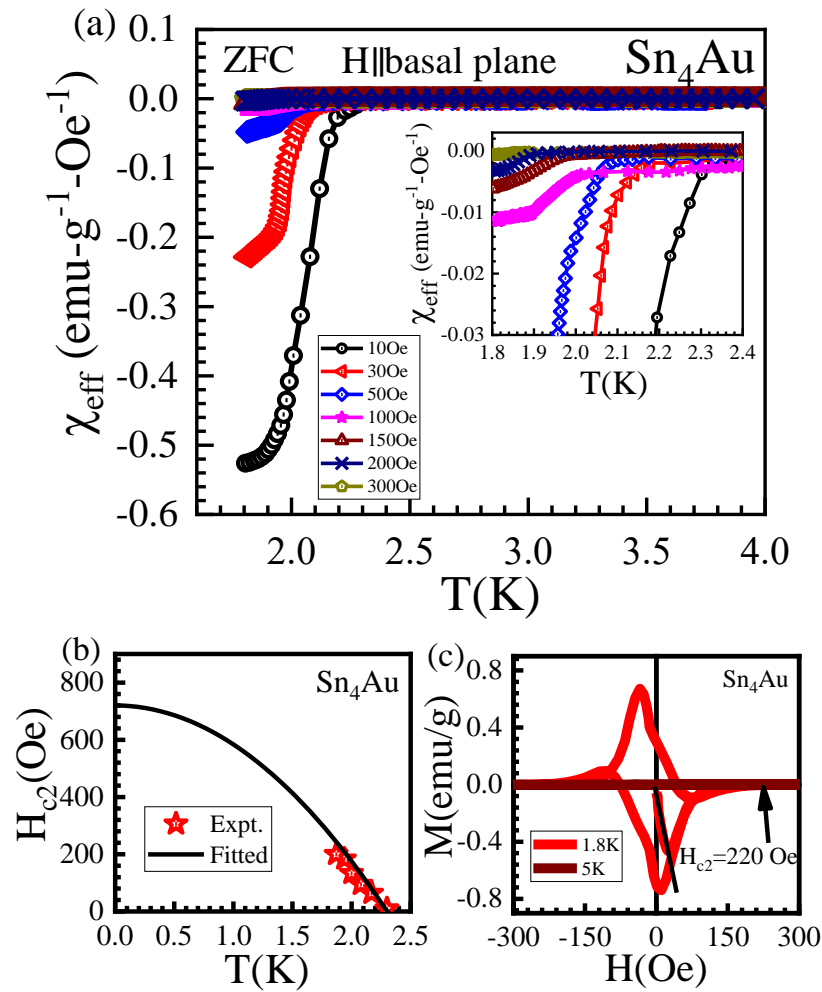

**Fig.2:(a)** DC magnetization (M-T) plots under ZFC protocol at different applied magnetic fields viz. 10, 30, 50, 100, 150, 200 and 300 Oe. (b) Fitted upper critical field vs temperature plot of synthesized $Sn_4Au$ single crystal. (c) Isothermal M-H plots of synthesized $Sn_4Au$ single crystal at 2 K and 5 K.

Resistivity vs. temperature (ρ-T) measurements are depicted in fig. 1(c), showing the presence of superconducting transition with $T_c^{onset}$ at around 2.3 ± 0.1 K. Zero resistivity is found at 2.1 ± 0.1 K, showing the transition width of about 0.2 K. This narrow transition width shows the purity of the sample. Further, DC susceptibility vs. temperature (M-T) measurements are carried out under zero field cooled (ZFC) protocols. Fig. 2(a) shows the susceptibility vs. T ($\chi_{eff}$ vs. T) plots at different DC magnetic fields viz. 10 Oe, 30 Oe, 50 Oe, 100 Oe, 150 Oe, 200 Oe, and 300 Oe. Magnetic measurements are performed on mechanically cleaved rectangular flake and the field is applied along the basal plane. The demagnetization factor is also calculated for the rectangular geometry of the sample using the following formula [30]

$$N = 1 - \frac{1}{1+\frac{q*a}{b}} \tag{1}$$

Here, a and b are dimensions of the sample in perpendicular and parallel to the magnetic field respectively, and these are 3.4 ± 0.1 mm and 0.16 ± 0.1 mm respectively. q represents the shape-dependent parameter, and for a rectangular sample, it is determined by the following formula

$$q = \frac{\pi}{4} + 0.64 tanh\left[0.64\frac{b}{a} ln\left(1.7 + 1.2 * \frac{a}{b}\right)\right] \tag{2}$$

The value of q is found to be 0.79, and the corresponding value of the demagnetization factor is N=0.943. The demagnetization factor is used to determine effective DC susceptibility with the relation $\chi_{eff} = \frac{\chi_m}{1+N*\chi_m}$, where $\chi_m$ is measured DC susceptibility. The superconducting volume fraction is found to be around 55 % at 10 Oe, showing the presence of bulk superconductivity in synthesized $Sn_4Au$ single crystal. The observed value of superconducting volume fraction is comparable to some of other superconducting topological materials [23,31,32]. The inset of fig. 2(a) depicts the zoom view of $\chi_{eff}$ - T plot around the superconducting transition. It is clear from the inset of fig. 2(a), the superconductivity persists up to 200 Oe and disappears at 300 Oe in synthesized $Sn_4Au$ single crystal. The upper critical field $H_{c2}$ is plotted against the temperature in fig. 2(b), and fitted with conventional superconductivity equation given as follows:

$$H_{c2}(T) = H_{c2}(0) * \left(1 - \frac{T^2}{T_c^2}\right) \tag{3}$$

Here, $H_{c2}(0)$ is defined as the upper critical field at absolute zero. $H_{c2}(0)$ is found to be 720 ± 1 Oe. Fig. 2(c) shows the M-H plot of synthesized $Sn_4Au$ single crystal at 2 K. M-H plot shows the presence of two distinct superconductivity critical fields viz. $H_{c1}$ and $H_{c2}$. $H_{c1}$ represents the lower critical field and is defined as the magnetic field at which the first magnetic line of force penetrates the superconducting sample. $H_{c1}$ is the point at which the low field M-H plot deviates from linearity, and the same is found to be around 35 ± 1 Oe. $H_{c2}$ is the field at which the M-H plot touches the baseline, and the same is found to be around 220 ± 1 Oe, which agrees with the $\chi_{eff}$ - T measurements.

The superconducting state of candidates of topological superconductivity is often characterized by two-fold anisotropic properties [20-24]. Here, angle-dependent magnetotransport measurements have been carried out to check the same. The synthesized crystals are grown unidirectionally and enable us to measure the angle-dependent properties of the same. $Sn_4Au$ is considered to be crystallized in Aea2 space group symmetry based on XRD results. Here, the current is passed along the (200) axis, and the magnetic field is applied along the (002) axis. The tilt angle (θ) is the angle between the applied magnetic field and the basal plane or the current direction. The schematic of the experimental setup is shown in fig. 3(c). Fig. 3(a) and 3(b) represent the temperature dependency of the upper critical field, i.e., $H_{c2}$, as obtained from magneto transport (ρ-H) measurements. The ρ-H measurements have been carried out along three characteristic directions, viz. in-plane direction (θ=0°) and out-of-plane directions (θ=90°). $H_{c2}$ is the point at which the resistivity rises to 50% of the normal state resistivity. $H_{c2}$ shifts to lower values as the temperature is increased. Thus, created phase diagram in all measured directions is shown in fig. 3(d).

A clear anisotropy in $H_{c2}$ is observed at temperatures below $T_c$ (2.3 K). This anisotropy is smaller at temperatures near to $T_c$ and increases as the temperature is further lowered. Anisotropy is also evident from the transition width, i.e., ΔH at 1.8 K, which is taken as the difference between $H_c^{onset}$ and $H_c^{offset}$. $\Delta H$ is found to be 108 ± 1 Oe, for $\theta=0°$, i.e., for in-plane measurements, and the same is found to be 150 ± 1 Oe for $\theta=90°$ out of plane measurements. The upper critical field at absolute zero, i.e., $H_{c2}(0)$ has been calculated for all three characteristic directions by fitting the $H_{c2}$ vs. T plots with equation 1. $H_{c2}^{\perp I}(0)$ is found to be 770 ± 1 Oe, nearly matching the obtained value from magnetization measurements. $H_{c2}^{\parallel I}(0)$ is found to be 970 ± 1 Oe, which is higher than the in-plane value, showing the anisotropic character of the upper critical field. The superconducting characteristic lengths are also derived from the upper critical field for different orientations of crystal. Expressions used for the calculation of superconducting coherence lengths are as follows: $H_{c2}^{\perp I}(0) = \frac{\Phi_0}{2\pi\xi_{\parallel I}^2}$, $H_{c2}^{\parallel I}(0) = \frac{\Phi_0}{2\pi\xi_{\parallel I}\xi_{\perp I}}$. Here, $\phi_0$ represents flux quanta, and its value is $2.07\times10^{-7}$ Oe-cm$^2$. The obtained values of $\xi_{\parallel I}$ and $\xi_{\perp I}$ are found to be 58 ± 1 nm and 74 ± 1 nm, respectively. The superconducting anisotropic ratio ($\Gamma$) is also calculated using the relation $\Gamma = \frac{H_{c2}^{\perp I}(0)}{H_{c2}^{\parallel I}(0)}$, and the same is found to be 1.26.

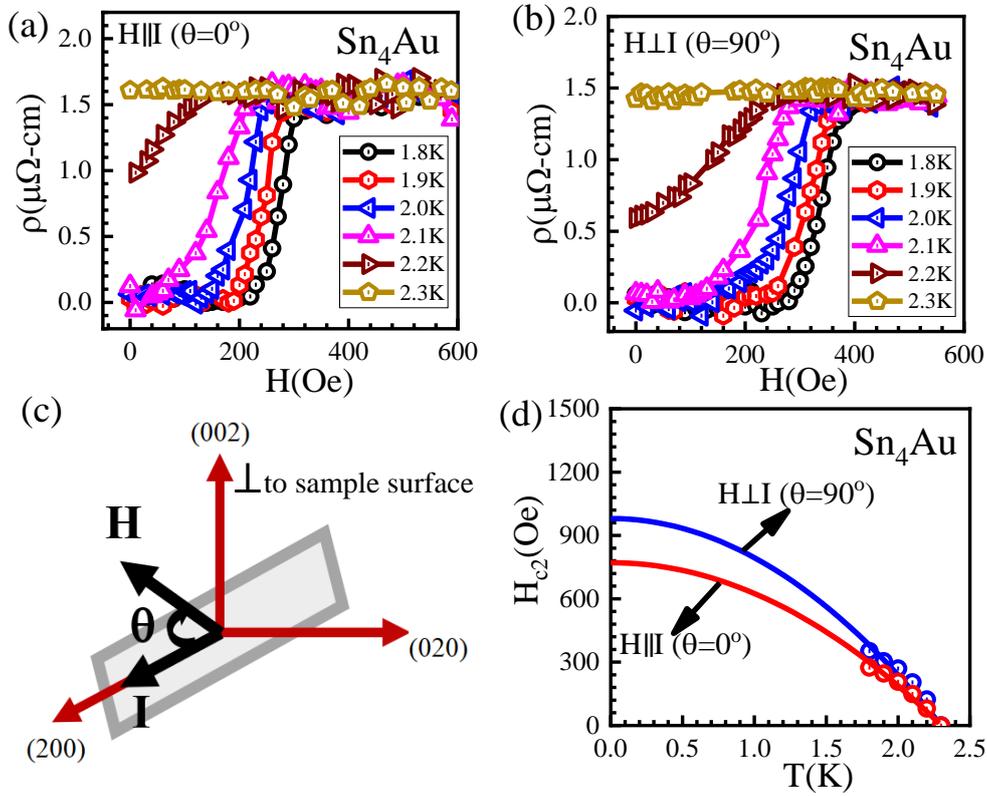

**Fig. 3:** $\rho$-H plots at different temperatures viz. 1.8, 1.9, 2.0. 2.1, 2.2 and 2.3 K at different field orientations as (a) $\theta=0°$ (b) $\theta=90°$ (c) Schematic of the geometry used for angle dependent magneto transport measurements. (d) Fitted upper critical field vs temperature plots at different field orientations.

To further elucidate the anisotropic properties of the upper critical field of synthesized Sn$_4$Au single crystal, $\rho$-H measurements have been performed at different orientations of crystal with respect to the applied field from 0º to 360º. The temperature is kept constant at 1.8 K (the lowest instrument limit) throughout the measurement. Fig. 4(a), (b), (c), and (d) are showing the $\rho$-H plots at different angle ranges viz. 0º-90º, 90º-180º, 180º-270º and 270º-360º respectively.

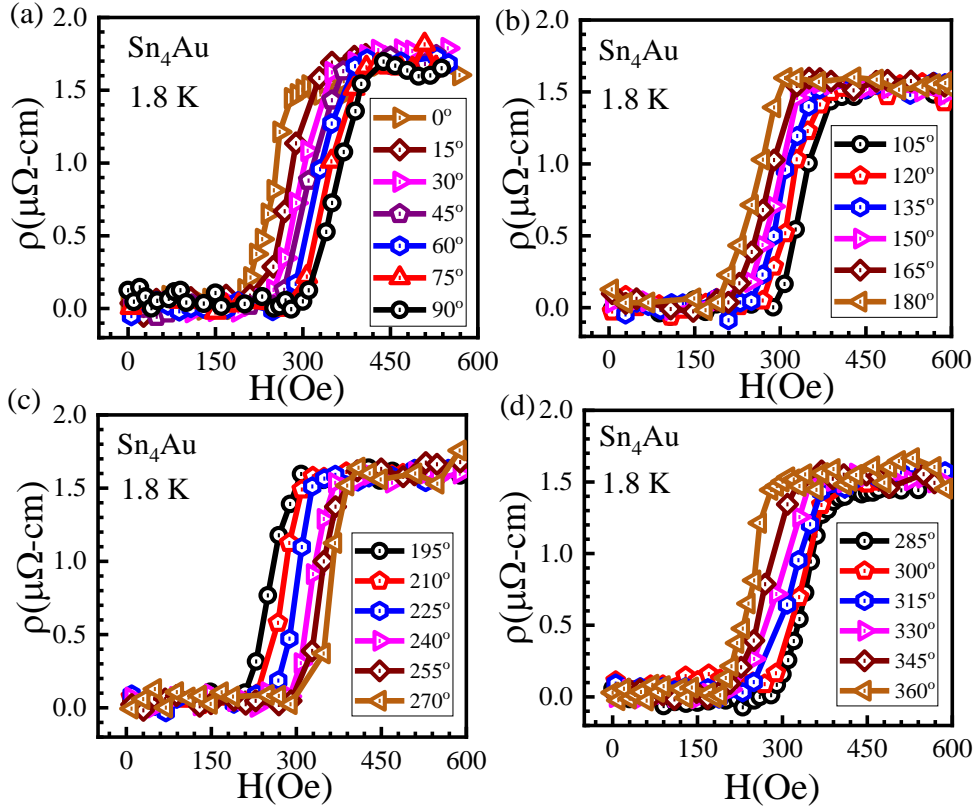

**Fig. 4:** ρ-H plots in different ranges of tilt angle θ viz. (a) θ=0º-90º (b) θ=105º-180º (c) θ=195º-270º (d) θ=285º-360º.

A clear shift in $H_c$ is observed as the orientation of the crystal with respect to the applied magnetic field changes. Here $H_c$ and $H_{irr}$ (irreversible field) are calculated for each orientation. $H_c$ is taken as the field at which the resistivity is 90% of the normal state resistivity, and the $H_{irr}$ is taken as the field at which the resistivity is 50% of its normal state value. Both the $H_{c2}$ and $H_{irr}$ are plotted against the orientation angle (θ) and are shown in fig. 5(a) and 5(c) respectively. Both plots have a sine wave-like appearance showing two-fold anisotropy of the upper critical field and irreversible fields. Fig. 5(b) shows the polar plot of $H_{c2}$ vs θ. Minima are observed for 0º, and maxima are observed for ±90º, indicating the upper critical field to be two-fold anisotropic. The two-fold symmetry can be well explained through Ginzberg-Landau (G-L) theory. According to G-L theory, anisotropy in effective mass leads to anisotropy in the upper critical field and can be given by the following equation

$$H_{c2}(\theta) = \frac{H_{c2}(0°)}{\sqrt{(cos^2\theta + \Gamma^{-2}sin^2\theta)}} \qquad (4)$$

Here $H_{c2}(0°)$ represents the upper critical field at 0º or along the basal plane, $\Gamma$ is an anisotropy parameter, and is found to be 1.37, which is near to as obtained from the ratio of $H_{c2}$ in different orientations. The $H_{c2}$ vs. θ plot is well fitted with equation (4), confirming the presence of a two-fold anisotropic superconducting state in $Sn_4Au$ single crystal.

$Sn_4Au$ lacks theoretical and experimental reports that could explain the observed two-fold anisotropic superconducting properties in the same. Topological superconducting materials are reported to show both in and out-of-plane anisotropic superconducting state due to breaking crystalline symmetry in the same [3,20-24,33,34].

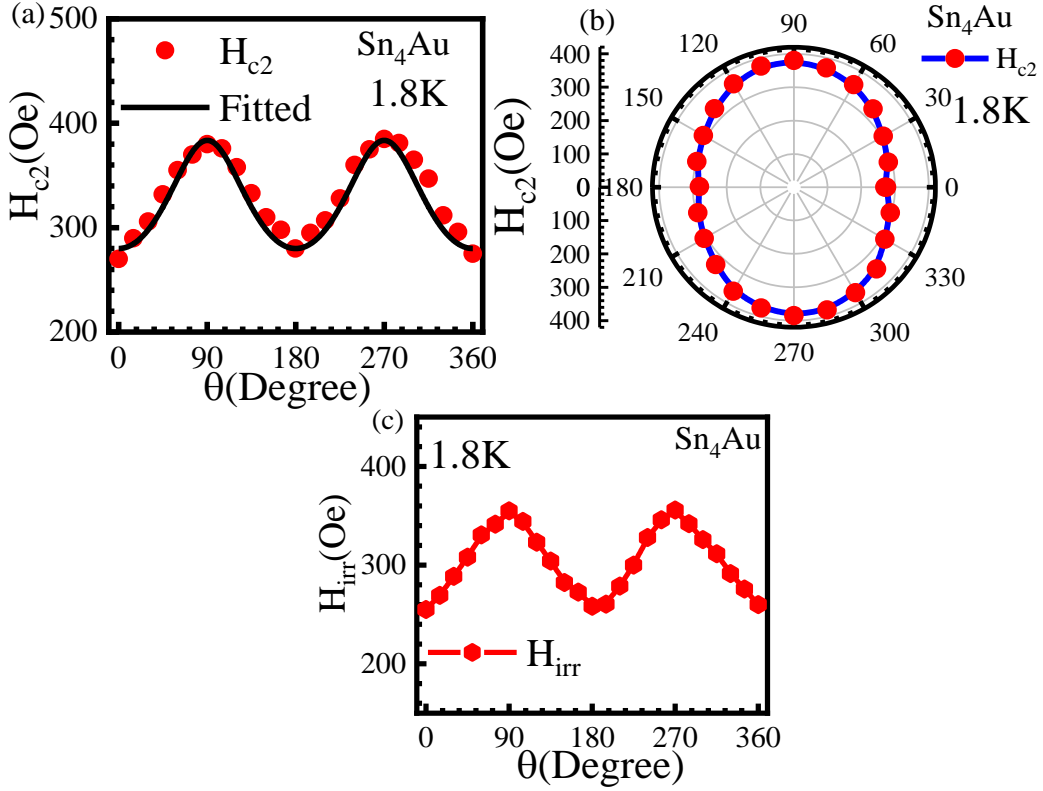

**Fig. 5:** **(a)** Variation of upper critical field $H_{c2}$ with tilt angle at constant temperature 1.8 K, which is fitted with G-L anisotropy equation. (b) Polar plot of $H_{c2}$ vs θ, showing the presence of two-fold anisotropy in the superconducting state. (c) $H_{irr}$ vs θ plot exhibiting sine wave like feature confirming the presence of two-fold anisotropy.

The other layered superconductors [35] are also found to show a similar feature as well. Here we observed the out-of-plane anisotropy in the superconducting state of $Sn_4Au$, by rotating the applied magnetic field with respect to the applied current. The observed anisotropic superconducting state in $Sn_4Au$, can be emerged due to layered structure of the same. The present results are important in the way that $Sn_4Au$, being crystallized in Aea2 space group symmetry, has a layered non-centrosymmetric structure. The non-centrosymmetric layered superconducting materials, with the absence of an inversion centre, are more suited to have a Rashba-Dresselhaus spin-orbit coupling effect [35-38]. The presence of the Rashba-Dresselhaus spin-orbit coupling effect leads to the presence of mixed superconducting pairing symmetries [36-38]. A similar example is layered $BiS_2$-based superconductors, which involve the breaking of inversion symmetry in a superconducting state and show both in-plane and out-of-plane anisotropic superconducting state [36,39]. The presence of Rashba-Dresselhaus spin-orbit coupling in $BiS_2$-bilayers leads to mixing of singlet and triplet superconducting states [39]. Very recently, some of the layered materials, namely monolayers of $NbSe_2$ [40], $TaS_2$ [41] are also reported to show in-plane two-fold anisotropic upper critical field due to the breaking of rotational symmetry. The synthesized $Sn_4Au$ crystal qualifies in both fronts as being layered topological material, and the same can be an excellent candidate to probe in-plane anisotropy in the superconducting state to get more insight in terms of its pairing symmetry and breaking of crystalline symmetry. More advanced measurements, such as high-resolution angle-resolved photoelectron spectroscopy (ARPES) or μ-SR measurements, can be fruitful for a more thorough study of the superconducting state of $Sn_4Au$.

**Conclusion:** Summarily, we investigated the angle-dependent magneto transport properties of as-grown $Sn_4Au$ single crystals. Superconductivity is evidenced by magnetization and transport measurement. In both resistivity and upper critical field measurements, a large anisotropy has been observed in the superconducting state. $H_{c2}$ is found to be higher in transverse orientation than the longitudinal one, which neglects the possibility of flux flow-induced two-fold anisotropy in the system. The observed two-fold anisotropic superconducting state seems to be appeared due to layered structure non-centrosymmetric structure of $Sn_4Au$. We hope that the present results will prove to be fruitful for further investigation of superconducting state properties of topological superconductor $Sn_4Au$.

**Acknowledgment:** The authors would like to thank Director NPL for his keen interest and encouragement. M.M. Sharma would like to thank CSIR, India, and AcSIR for the research fellowship and Ph.D. registration respectively.